\documentclass[aps,prl,
 reprint,
 amsmath,amssymb,
 superscriptaddress]{revtex4-1}

\usepackage{graphicx}
\usepackage{amsmath}
\usepackage{braket}
\usepackage{dcolumn}
\usepackage{bbm}
\usepackage{siunitx}
\usepackage[colorlinks=true,allcolors=blue,bookmarks=false,pdfusetitle]{hyperref}

\usepackage[caption=false]{subfig}
\begin{document}

\title{A strongly interacting photonic quantum walk using single atom beam splitters}

\author{Xinyuan Zheng}
 \email{xzheng16@terpmail.umd.edu}
 \affiliation{Department of Electrical and Computer Engineering and Institute for Research in Electronics and Applied Physics, University of Maryland, College Park, Maryland 20742, USA}
\author{Edo Waks}%
 \email{edowaks@umd.edu}
 \affiliation{Department of Electrical and Computer Engineering and Institute for Research in Electronics and Applied Physics, University of Maryland, College Park, Maryland 20742, USA}
 \affiliation{Joint Quantum Institute, University of Maryland, College Park, Maryland 20742, USA}
 \affiliation{Department of Physics, University of Maryland, College Park, Maryland 20742, USA}

\date{\today}

\begin{abstract}

Photonics provide an efficient way to implement quantum walks, the quantum analogue of classical random walk that demonstrates rich physics with potential applications. However, most photonic quantum walks do not involve photon interactions, which limits their potential to explore strongly-correlated many-body physics of light. We propose a strongly interacting discrete-time photonic quantum walk using a network of single atom beamsplitters. We calculate output statistics of the quantum walk for the case of two photons, which reveals the strongly-correlated transport of photons. Particularly, the walk can exhibit either boson-like or fermion-like statistics which is tunable by post-selecting the two-photon detection time interval. Also, the walk can sort different types of two-photon bound states into distinct pairs of output ports under certain conditions. These unique phenomena show that our quantum walk is an intriguing platform to explore strongly-correlated quantum many-body states of light. Finally, we propose an experimental realization based on time-multiplexed synthetic dimensions.

\end{abstract}
\maketitle



Photonics provides an efficient way to implement quantum walks\cite{Childs2003,Shenvi2003,Childs2009,Childs2013a,Kempe2003,Qiang2021,Gong2021}, which is the quantum analogue of classical random walk. Photonic quantum walks are essentially a family of transport phenomena of one or few photons in a given structure, and have been used to demonstrate rich physics and to simulate Hamiltonian evolutions. Photonic quantum walks have been proposed and implemented in discrete-time via platforms such as beamsplitter arrays\cite{Sansoni2012,Schreiber2010,Kitagawa2010,Schreiber2012a} or in continuous-time via platforms such as coupled waveguides\cite{Poulios2014a,Tang2018b}. However, the majority of these examples utilize linear optical elements and cannot induce photon-photon interactions during the walk. In contrast, quantum walks with strong multi-particle interactions enables the study of strongly correlated many-body physics and have attracted significant interest\cite{Yan2019,Preiss2015}.

In continuous-time, circuit QED enables a strongly interacting quantum walk of microwave photons\cite{Yan2019}. Even for two microwave photons, the walk has given rise to many fundamentally novel quantum many-body transport effects, such as the variation of statistics with respect to interaction strength, or fermionization\cite{Preiss2015,Yan2019}. In discrete time, however, photonic quantum walks are mostly proposed and implemented using linear beam splitters, which cannot induce photon-photon interaction. A natural nonlinear substitute of the linear beam splitter is the single atom beamsplitter, consisting of an atom strongly coupled to a 1D waveguide\cite{Shen2005,Shen2007a}. 
Although this system acts like a linear beamsplitter for a single quasi-monochromatic input photon\cite{Shen2005}, it can induce strong photon-photon interactions during the multiphoton scattering process\cite{Shen2007a,Xu2015,Mahmoodian2020} and generate strongly correlated multiphoton output states such as photon bound states. 
Recently, it is shown that this nonlinearity can induce strongly correlated multiphoton transport in simpler structures such as a 1D chiral chain\cite{Mahmoodian2020,Iversen2021,Javadi2015,Goban2014,Prasad2020,Yang2022,Tiecke2014}. However, leveraging this beamsplitter in a discrete-time photonic quantum walk is largely unexplored and poorly understood due to its complexity.

%

Here we propose and study a strongly interacting discrete-time photonic quantum walk using single atom beamsplitters. We study the dynamics of the walk for the case of two photons, and obtain the complete two-photon statistics at the output. We show that the walk can exhibit either boson-like or fermion-like statistics which is tunable by post-selecting the two-photon detection time interval. This feature is an interesting analogue to the continuous-time walk with strong interaction, whose statistics is tuned by the interaction strength $U$\cite{Yan2019,Preiss2015}. Moreover, the walk can sort different types of two-photon bound states into distinct pairs of output channels under certain conditions. These fundamental quantum effects show the clear signature of a strongly interacting photonic quantum walk and provide the discrete-time analog to the continuous-time model\cite{Yan2019,Preiss2015}. Finally, we propose a practical implementation of the discrete-time random walk based on time-multiplexed synthetic dimensions\cite{Schreiber2010,Schreiber2011,Schreiber2012a,Nitsche2016}, which provides a viable pathway towards experimental realization.

\begin{figure}[t]
\subfloat[\label{fig:1a}]{}
\subfloat[\label{fig:1b}]{}
\subfloat[\label{fig:1c}]{}
\subfloat[\label{fig:1d}]{}
\includegraphics[width=3.4in]{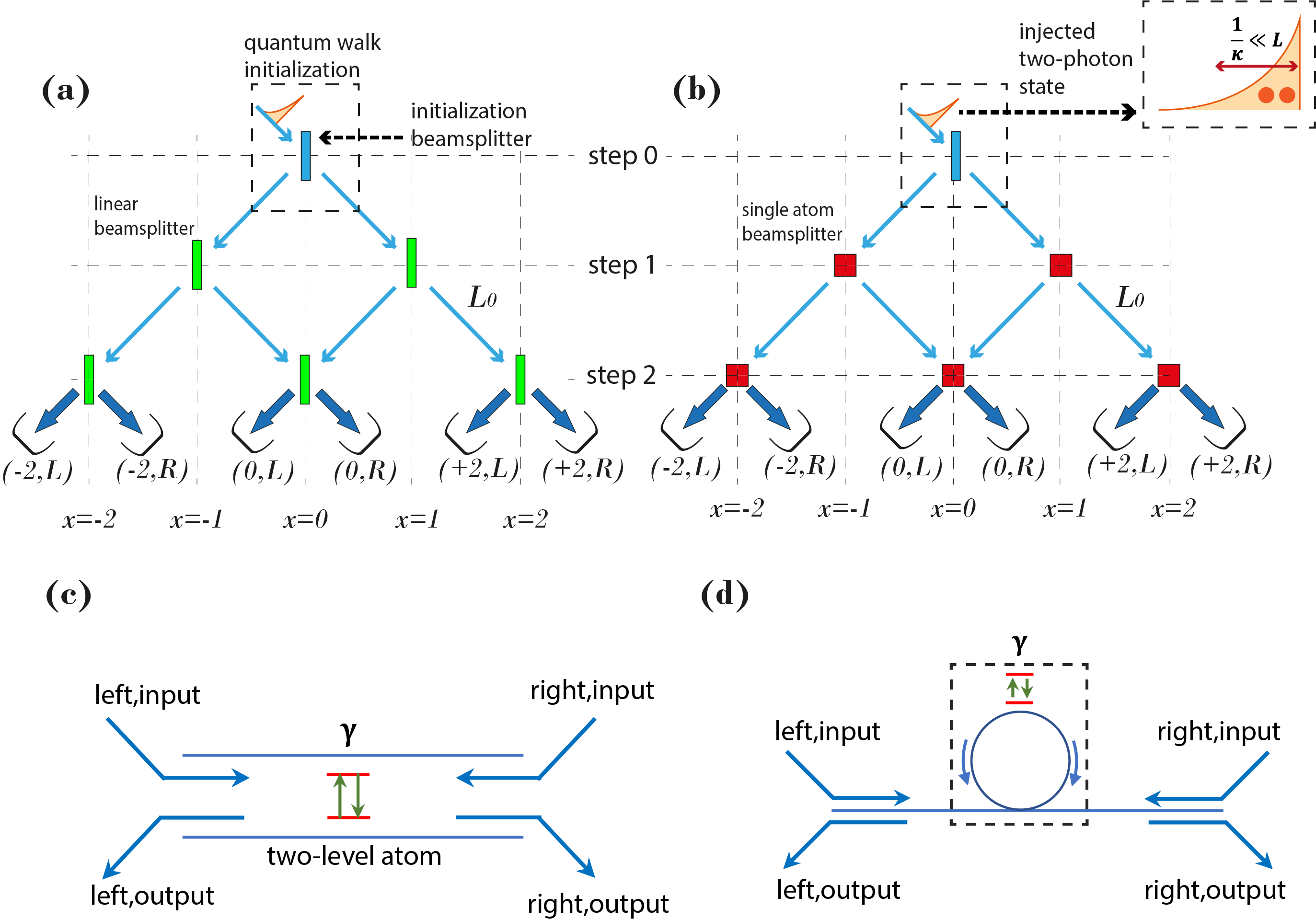}
\caption{\label{fig:1}(a)A three-step linear quantum walk. At each step $n\geq1$, the linear beamsplitters partition each walker to the left or right. The square brackets represent photon counters that detect the final positions and directions of the output photons. (b)The nonlinear Galton board consists of an array of single atom beamsplitters. We use the same source at step $0$ to generate the initial state of the walk (c)A single atom beamsplitter composed of a two-level system coupled to a waveguide. (d)An alternative implementation using an atom-cavity system in the fast cavity regime\cite{Rosenblum2011a,Dayan2008} in place of the two-level system in (c).}
\end{figure}

Our nonlinear random walk is based on the architecture of the discerete-time linear quantum walk shown in Figure~\ref{fig:1a}.
This architecture consists of an array of linear beamsplitters interconnected by feedforward waveguides, forming a quantum Galton board. 
At step $0$, we initialize the walk by injecting a few-photon state towards a tunable initialization beamsplitter, which is described by a $2\times2$ unitary transform matrix acting on the two input modes and produces the desired output state that propagates to step $1$. The spatial separation between adjacent steps is denoted $L_0$, which is taken to be much larger than the pulse width of the injected light. At each step $n\geq1$, the identical beamsplitters implement a local, unitary transformation to the incoming input state and generate the outgoing output state propagating to the next step. Finally, photodetectors located after the final step $(N)$ are used to measure the output states, yielding the output statistics of the walk. We label each detector with a pair of index $(x,d)$, where $x$ denotes the horizontal location and $d\in{L,R}$ is the direction of output. The linear walk has been used to demonstrate a variety of multiphoton interference phenomena for different few-photon initial states\cite{Sansoni2012}, but the linear beamsplitters cannot induce photon-photon interactions.

To introduce a random walk with strong interactions, we replace the linear beamsplitter in Figure~\ref{fig:1a} with single atom beamsplitters as illustrated in Figure~\ref{fig:1b}. Each single atom beamsplitter consists of a two-level atom coupled to the left(right)-propagating waveguide modes, with two inputs and two outputs(Figure~\ref{fig:1c}). An alternative implementation uses an atom-cavity system where the cavity is in the fast cavity regime(Figure~\ref{fig:1d}). The atom has resonant frequency $\Omega$ and an effective decay rate $\gamma$ into the waveguide, while the group velocity of waveguide photon is set to $v_g=1$. Each atom interact with waveguide photons via a local, Hermitian Hamiltonian $H_{int}\propto (b_{+}(y=nL_{0})a^\dagger_{n,x}+b_{-}(y=nL_{0})a^\dagger_{n,x}+h.c.)$, where $a_{n,x}$ is the atomic operator at step $n$, location $x$ and $b_{+(-)}(y=nL_{0})$ annihilates a right(left) propagating photon at $y=nL_{0}$, which is where atom $(n,x)$ is coupled to the waveguide. Hence, any photon state injected towards the single atom beamsplitter undergoes a local unitary evolution at the proximity of the atom before it fully leaves the beamsplitter. We may derive an input-output relation for each beamsplitter based on the interacting Hamiltonian(see supplement):
\begin{equation}
\begin{aligned}
b_{n,x;R(L),out}(t) &= b_{n,x;L(R),in}(t) - i\sqrt{\gamma}a_{n,x}(t)\\
\end{aligned}
\label{equ:1}
\end{equation}
Here $b_{n,x;L(R),in(out)}(t)$ is the input(output) operator on the left(right). 
The dynamics of the atomic operator is given by\cite{Xu2013,Xu2015}:
\begin{equation}
\begin{aligned}
\frac{da_{n,x}}{dt} = -i\sqrt{\gamma}&[a_{n,x},a^{\dagger}_{n,x}](b_{n,x;L,in}+b_{n,x;R,in})-\\
i&[a_{n,x},H_{atom}-i\gamma a_{n,x}^\dagger a_{n,x}]\\
\end{aligned}
\label{equ:2}
\end{equation}  
Similar to the linear beamsplitter, the single atom beamsplitter is equipped with a scattering matrix $\hat{S}$, which can be found directly by solving equation~\ref{equ:1} or the Hamiltonian\cite{Shen2005,Shen2007a}. 
For a quasi-monochromatic left-incoming single-photon state $\ket{k,L}$ with frequency $k$, the $\hat{S}$-matrix elements are given by $\bra{p,L}\hat{S}\ket{k,L}=r_k\delta(k-p)$ and $\bra{p,R}\hat{S}\ket{k,L}=t_k\delta(k-p)$. For such an input state, the single atom beamsplitter therefore acts as a linear beamsplitter with transmission and reflection coefficients given by $t_k =\frac{k-\Omega}{i\gamma+(k-\Omega)}$ and $r_k =\frac{-i\gamma}{i\gamma+(k-\Omega)}$\cite{Shen2005}.
However, for two-photon input state, the $\hat{S}$-matrix elements are significantly more complex. For example, for $LL\rightarrow LL$ scattering\cite{Shen2007a,Shen2015a,Xu2013,Xu2015}:
\begin{equation}
\begin{aligned}
&\bra{p_1,p_2;L,L}\hat{S}\ket{k_1,k_2;L,L}=r_{k_1}r_{k_2}[\delta(k_1-p_1)\delta(k_2-p_2)\\
&+\delta(k_1-p_2)\delta(k_2-p_1)]+B(k_1,k_2,p_1,p_2)\delta(k_1+k_2-p_1-p_2)
\end{aligned}
\end{equation} 
The first two terms are noninteracting terms identical to a dispersive linear beamsplitter $\hat{S}$-matrix for two photons, but the third term is the nonlinear interacting term that implies strong photon-photon interactions\cite{Shen2015a,Shen2007a,Xu2013}.
To initialize the walk, at $T=0$ we inject a two-photon pulse $\ket{ini}=\frac{1}{\sqrt{2}}\int dy_1dy_2 \psi(y_1,y_2)b^{\dagger}(y_1)b^{\dagger}(y_2)\ket{vac}$ from the left towards the initialization beamsplitter at step $0$ as shown in Figure~\ref{fig:1b}.
The input-output relation for the initialization beamsplitter is chosen to be $b_{0,0;L,out}(t)=[b_{0,0;L,in}(t)-b_{0,0;R,in}(t)]/\sqrt{2}$ and $b_{0,0;R,out}(t)=[b_{0,0;L,in}(t)+b_{0,0;R,in}(t)]/\sqrt{2}$. The initialization beamsplitter therefore outputs a pair of indistinguiable photon where each photon has $50:50$ probability heading $L/R$. Although we can numerically solve the walk for any $\psi(y_1,y_2)$, we consider specifically an exponential pulse
\begin{equation}
\psi(y_1,y_2)\propto
\begin{cases}
e^{(i\omega+\kappa)(y_1+y_2)} & y_1,y_2<0 \\
0 & otherwise
\end{cases}
\end{equation}
This allow us to solve for the output analytically with an efficient algorithm(see supplement). We therefore focus on this special case.
Furthermore, we can freely control the parameter $\kappa$, which determines the bandwidth of each photon and the parameter $\omega$ which determines each photon’s detuning $\delta=\omega-\Omega$ with respect to the atom’s resonant frequency.

We perform our calculations for initial state satisfying $\kappa\ll\gamma$ and $\frac{1}{\kappa}\ll L_0$.
The first condition states that the injected photons have a narrow spectral bandwidth compared to the atomic transition, hence the noninteracting term in the single atom beamsplitter $\hat{S}$ matrix can be approximated by that of a linear beamsplitter with reflection(transmission) coefficient $r_\omega$ and $t_\omega$. We can therefore extract the effect of interacting term by comparing our results with a linear walk consisting of those beamsplitters. The second condition guarantees that the photon’s pulse width is much less than the traveling distance between adjacent steps, and hence the photons are synchronized within the same step. They therefore both arrive at step $n$ at time $nL_0$, undergo the unitary scattering process at that step, before fully leaving step $n$ and arriving at step $(n+1)$ at time $(n+1)L_0$.
After an $N$-step walk, we measure the output state of the walk with $2N+2$ photodetectors(\ref{fig:1b}) and obtain a time-ordered two-photon statistics. 
The full statistics is given by:
\begin{equation}
\begin{aligned}
&\Gamma_{x_2,d_2;x_1,d_1}(t,\tau)=\\
&\langle b^{\dagger}_{N,x_1;d_1,out}(t+NL_0)b^{\dagger}_{N,x_2;d_2,out}(t+\tau+NL_0)\\
&b_{N,x_2;d_2,out}(t+\tau+NL_0)b_{N,x_1;d_1,out}(t+NL_0) \rangle
\end{aligned}
\end{equation}
Physically, the quantity represents the time-ordered probability density of detecting the first photon at time $t+NL_0$ in detector $(x_1,d_1)$ and the second at time $t+\tau+NL_0$ in detector $(x_2,d_2)$.
To observe the effect of nonlinearity, we follow the approach in most previous works and compute the second order correlations, given by:
\begin{equation}
G_{x_1,d_1;x_2,d_2}(\tau)=\int_{0}^{\infty}dt(\Gamma_{x_1,d_1;x_2,d_2}(t,\tau)+\Gamma_{x_2,d_2;x_1,d_1}(t,\tau))
\end{equation}
The second order correlation functions represent the probability density of detecting two photons separated by a delay $\tau$ in detector $(x_1,d_1)$ and $(x_2,d_2)$. We denote the matrix $G_{x_1,d_1;x_2,d_2}(\tau=\tau_0)$, with two indices $(x_i,d_i)$ running over all $(2N+2)$ detectors at step $N$, as the statistical pattern obtained by post-selecting all two-photon detections with a certain time interval $\tau_0$. We will show that $\tau_0$ plays an essential role in modifying the statistical pattern, analogous to the interaction strength $U$ in continuous-time walks\cite{Yan2019}. The complete details for solving these correlations are given in the supplement.
\begin{figure}[t]
\subfloat[\label{fig:2a}]{}
\subfloat[\label{fig:2b}]{}
\subfloat[\label{fig:2c}]{}
\subfloat[\label{fig:2d}]{}
\subfloat[\label{fig:2e}]{}
\subfloat[\label{fig:2f}]{}
\includegraphics[width=3.4in]{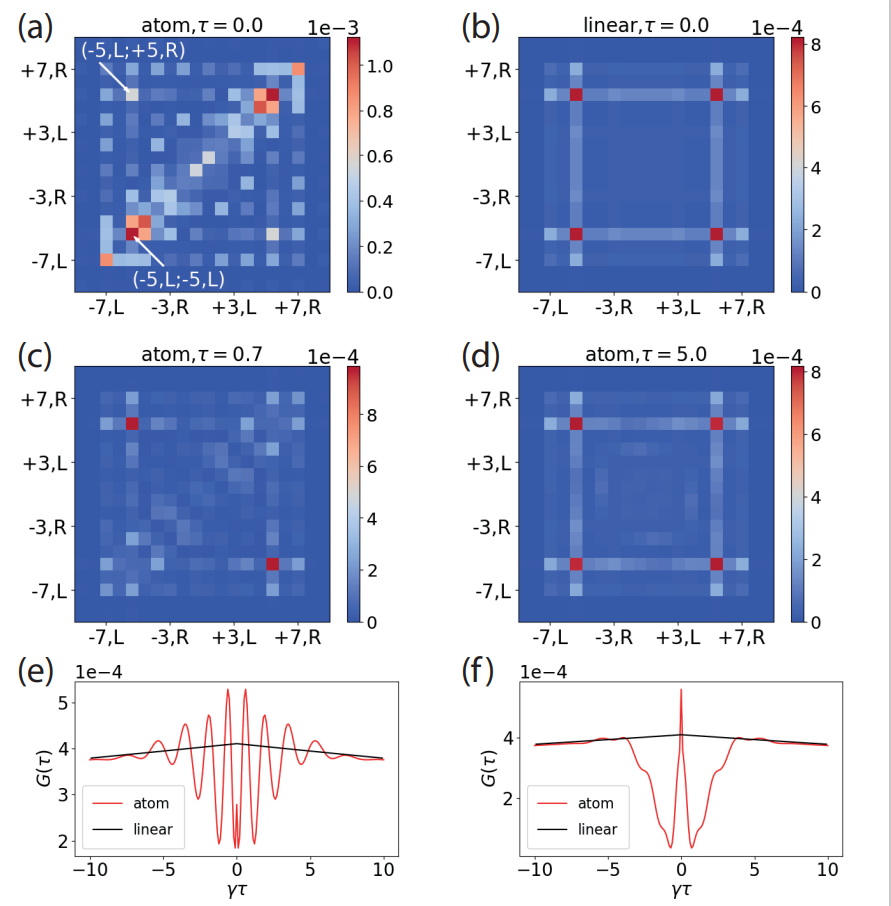}
\caption{\label{fig:2}Results for $\kappa\ll\delta=\gamma$.(a)The statistical pattern matrix for the nonlinear Galton board for $\tau=0$, showing correlated statistics. (b)The statistical pattern for $\tau=0$ for the linear Galton board. (c-d)The statistical pattern for the nonlinear Galton board for $\tau=0.7,5.0$, showing anti-correlated and uncorrelated statistics. (e)$G_{-5,L;+5,R}(\tau)$ and $G_{+5,R;-5,L}(-\tau)$, plotted for all $0\leq\tau\leq10$. (f) $G_{-5,L;-5,L}(|\tau|)$for all $0\leq\tau\leq10$.}
\end{figure}

We first analyze the case where $\gamma=\delta=1$ and $\kappa=0.002$. In this case, the noninteracting term in the $\hat{S}$-matrix can be approximated by that of a linear beamsplitter with $r=\frac{-i}{1+i}$ and $t=\frac{1}{1+i}$.
Figure~\ref{fig:2a} shows the statistical pattern of the nonlinear Galton board after 9 steps after post-selecting $\tau=0$. Here we have a boson-like statistical pattern where the photons are more likely to be found by the same detector.
In comparison, Figure~\ref{fig:2b} shows the result for the linear Galton board for $\tau=0$. Here the two walkers independently partition themselves to the left and right, showing no statistical correlations.
Interestingly, for a slightly larger $\tau=0.7$(Figure~\ref{fig:2c}), the statistical pattern for the interacting quantum walk flows from boson-like to fermion-like where the photons are more likely to be found by detectors on opposite sides of the Galton board. Namely, the post-selected detection interval $\tau$ serves as a tuning knob that tunes the output statistical pattern, similar to the interaction strength $U$ in continuous time walks\cite{Yan2019,Preiss2015}.
Notably, in contrast to the $\tau$-dependent statistical pattern for the nonlinear Galton board, the statistical pattern for the linear Galton board remains the same for any $\tau$(see supplementary section), which is identical to Figure~\ref{fig:2b}.
Finally, Figure~\ref{fig:2d} shows the case for $\tau=5.0$. Here, the post-selected photon detection events are sufficiently separated in time (i.e. $\tau\gg1/\gamma$) and the interaction is effectively turned off. We therefore have no discrepancy between the statistical pattern of the linear and nonlinear walk.

To attain a better understanding of the varying statistical pattern, we select two representative pairs of detectors, $\{-5,L;+5,R\}$ and $\{-5,L;-5,L\}$, as indicated in Figure~\ref{fig:2a}, and plot $G_{x_2,d_2;x_1,d_1}(\tau)$ against all $\tau$. For the pair of detector on opposite sides of the Galton board $\{-5,L;+5,R\}$, we observe an oscillation in $G(\tau)$ that gradually dies out for $\tau>6$ as shown in Figure~\ref{fig:2e}. For $\{-5,L;-5,L\}$, we observe a narrow peak around $\tau=0$ followed by a valley with minimal value at $\tau=0.7$(Figure~\ref{fig:2f}). In comparison, the correlation functions $G(\tau)$ for the linear Galton board are two identical smooth curves shown in black(Figure~\ref{fig:2e},\ref{fig:2f}), indicating no photon-photon interaction.

\begin{figure}[t]
\subfloat[\label{fig:3a}]{}
\subfloat[\label{fig:3b}]{}
\subfloat[\label{fig:3c}]{}
\subfloat[\label{fig:3d}]{}
\subfloat[\label{fig:3e}]{}
\subfloat[\label{fig:3f}]{}
\includegraphics[width=3.4in]{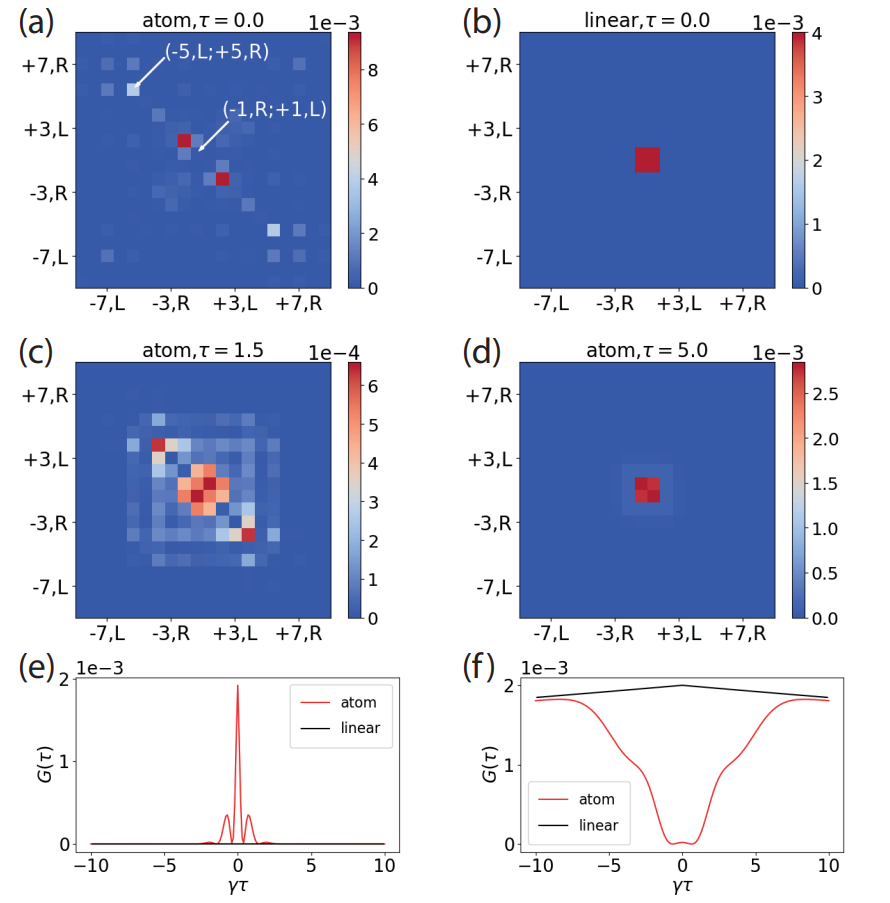}
\caption{\label{fig:3}Results for $\delta=0$, $\kappa\ll\gamma$.(a)The statistical pattern matrix for the nonlinear Galton board for $\tau=0$, showing anti-correlated statistics excluding the center. (b)The statistical pattern for $\tau=0$ for the linear Galton board. (c-d)The statistical pattern for the nonlinear Galton board for $\tau=1.5,5.0$, converging to the pattern in (b) as $\tau$ increases. (e)$G_{-5,L;+5,R}(\tau)$and $G_{+5,R;-5,L}(-\tau)$, plotted for all $0\leq\tau\leq10$. (f)$G_{-1,R;+1,L}(\tau)$ and $G_{+1,L;-1,R}(-\tau)$ for all $0\leq\tau\leq10$.}
\end{figure}

The nonlinear Galton board can also generate quantum walks with unequal splitting ratios by simply changing injected photon's detuning $\delta$ with respect to the atom's resonant frequency $\Omega$. As an illustrative example, we may choose $\delta\equiv0$, corresponding to on-resonant input photons. We therefore compare our results with that of a linear quantum walk with $100:0$ splitting ratio(i.e. $r=1$). Figure~\ref{fig:3a} shows the statistical pattern for $\tau=0$, while Figure~\ref{fig:3b} shows the result for the linear Galton board. 
For the linear Galton board, the photons are trapped at the center since they reflect back and forth repeatedly between adjacent beamsplitters around $(x=0)$. Surprisingly, the nonlinear Galton board exhibits a drastically different statistical pattern that predominantly excludes the center(Figure~\ref{fig:3a}). Instead, local maximums appear at $\{-x,L;+x,R\}$ for $x=1,3,5$, showing higher probability of detecting the two photons in these pairs of detectors. At $\tau=1.5$ (Figure~\ref{fig:3c}) this statistical pattern gradually shrinks towards the center, while at $\tau=5.0$ it becomes almost identical to the linear case(Figure~\ref{fig:3d}). 

To understand these pattern, we investigate $G(\tau)$ for $\{-x,L;+x,R\}$. For example, the correlation function $G(\tau)$ measured by detector-pair $\{-5,L;+5,R\}$ exhibits a high peak around $\tau=0$ as shown in Figure~\ref{fig:3e}, which is a typical signature for two-photon bound states. For $(x=1,3)$, we observe similar correlations, with slightly different $G(\tau)$ curves(see supplement). The Galton board is therefore sorting two-photon bound states out of these distinct pairs of outputs on the opposite sides of the Galton board. In contrast, for detector-pair $\{+1,L;-1,R\}$ at the center(Figure~\ref{fig:3f}), the photon correlation function exhibits a deep valley with minimum $0$ around $\tau=0$. Again, the $G(\tau)$ for the linear Galton board are smooth curves showing no sign of photon-photon interactions.

\begin{figure}[ht]
	\subfloat[\label{fig:4a}]{}
	\subfloat[\label{fig:4b}]{}
	\includegraphics[width=3.4in]{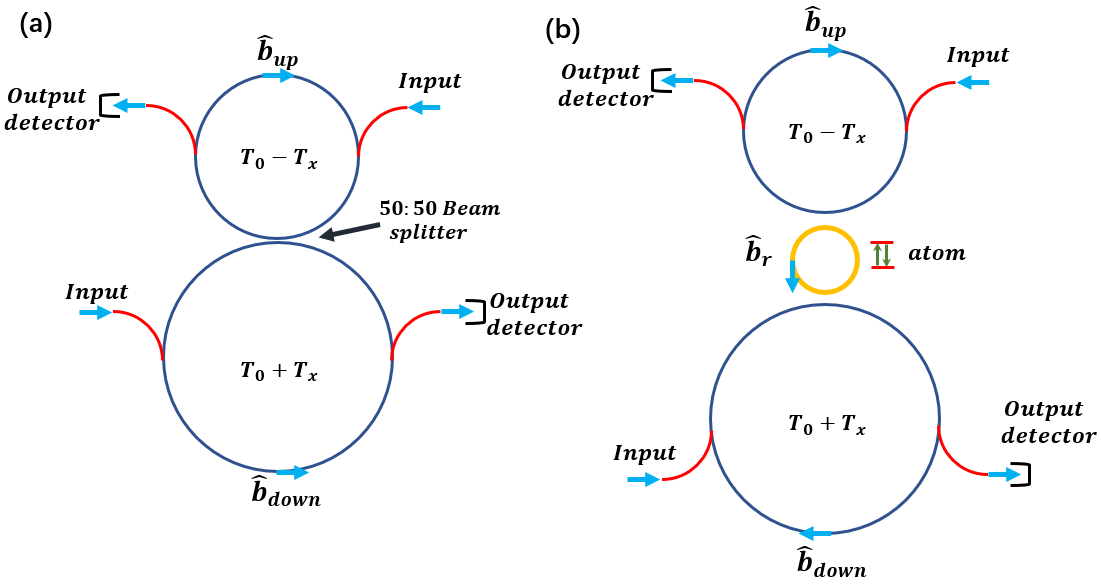}
	\caption{\label{fig:4}(a)Implementing a quantum walk with linear 50/50 beam splitters using time-multiplexed synthetic dimensions. (b) A nonlinear quantum walk implemented with a single atom beam splitter. The single atom beam splitter consists of a two-level atom chirally coupled to the $\hat{b}_{r}$ mode of a ring resonator(orange). In the "fast cavity regime"\cite{Rosenblum2011a}, the setup is equivalent to an atom coupled chirally to the two feedback loop modes, $\hat{b}_{up}$ and $\hat{b}_{down}$. }
\end{figure}

Having analyzed the properties of the nonlinear Galton board, we next consider potential experimental realizations.
We propose an implementation using time-multiplexed synthetic dimensions\cite{Nitsche2016,Schreiber2010,Schreiber2011,Schreiber2012a} to 
avoid the need to fabricate large arrays of single atom beam splitters. Figure~\ref{fig:4} shows an implementation of the linear Galton board using this method, which requires only one beam splitter coupled to two feedback loops with different time delays, $T_0+T_x$ and $T_0-T_x$. Photons are injected into the system via the two input ports and photon counters at the output ports detect the photons. The same beam splitter at different time delays $nT_0+xT_x$ acts as different beam splitters in the Galton board, provided that the pulse width $d$ of the photons satisfies $d\ll v_gT_x$ and $T_x\ll T_0$\cite{Schreiber2010}. All the photon counters in Figure~\ref{fig:1a} are represented by the two photon counters in Figure~\ref{fig:4a} at different time delays\cite{Schreiber2010}.

Taking advantage of this idea, we propose the system shown in Figure~\ref{fig:4b} to implement the nonlinear Galton board. 
The beam splitter consists of an atom chirally coupled to the ring resonator mode $\hat{b}_r$\cite{Sollner2015,Lodahl2017,Pichler2015,Mahmoodian2020}. The mode $\hat{b}_r$ is then directionally coupled to the upper loop mode $\hat{b}_{up}$ and lower loop mode $\hat{b}_{down}$. In the "fast cavity regime"\cite{Rosenblum2011a,Dayan2008}, one can adiabatically eliminate $\hat{b}_r$ and hence the atom is chirally coupled to $\hat{b}_{up}$ and $\hat{b}_{down}$ with some effective decay rate\cite{Rosenblum2011a}. The use of synthetic dimensions then enables us to build the nonlinear Galton board with this single device.

In summary, we have proposed and studied a strongly interacting photonic quantum walk using single atom beamsplitters. By post-selecting photon detection events with certain time intervals, we can tune the output statistics of a two-photon walk. Our quantum walk model can be easily generalized to more complicated quantum walks in higher dimensions or with nontrivial topology\cite{Kitagawa2012a,Kitagawa2012,Kitagawa2010,Kitagawa2010a,Barkhofen2017,Chen2018}. We may also extend this work by choosing other types of few-photon input states, such as the coherent state, squeezed states, and entangled states of light. Moreover, our formalism can be applied to a general family of light propagation problems consisting of feedforward networks with nonlinear input-output nodes, such as the quantum neural net\cite{Steinbrecher2019}. Ultimately, our work can be extended in many ways and opens the door of a whole class of models to study strongly-correlated many-body states of light.

\begin{acknowledgments}
The authors would like to acknowledge financial support from the National Science Foundation (grant number OMA1936314, OMA2120757 and ECCS1933546), the Air Force Office of Scientific Research (grant number UWSC12985 and FA23862014072), and the Army Research Laboratory (grant W911NF1920181). 
\end{acknowledgments}

\bibliography{first_paper_ref}
\end{document}


\section{Fundamentals}
In this section, we write down the complete input-output relations for the entire quantum walk following \cite{Xu2015,Caneva2015a}. 
As an illustrative example, a walk with total step number $N=2$ is shown below: 
\begin{figure}[hbt]
\includegraphics[width=3.9in]{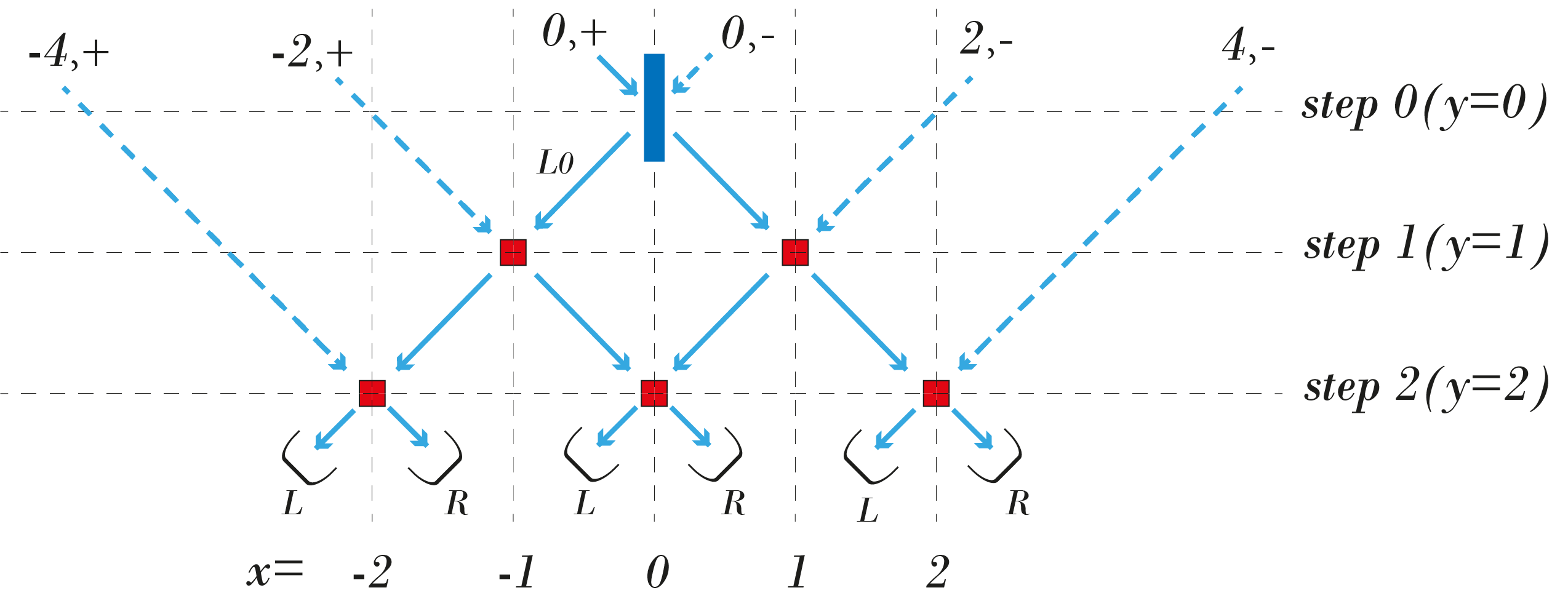}
\caption{A 2-step walk using 5 single atom beam splitters.}
\label{fig:s1}
\end{figure}

The system is described by a set of waveguide bosonic operators and atomic operators. Each string of blue arrow is actually representing a 
waveguide, and each waveguide is coupled to a string of beamsplitters. Each beamsplitter, either the "source beamsplitter"
or the single atom beamsplitter, offers a specific input-output relations at that node.
A waveguide intersecting step $0$ at $x=m$ and sending photon in the left(right) direction is labeled $(m,-(+))$. A waveguide labeled $(m,\pm)$ is coupled to a string of beamsplitter. For example, the waveguide labeled $(2,-)$ is coupled to the single atom beamsplitter at $(step=1,x=1)$ and $(step=2,x=0)$. For the bosonic field in each waveguide, we define:$b_{m,\nu}(y,t)$ for $y\in\mathbf{R}$ as the field operator. Since we are choosing Heisenberg picture, these operator depend on time $t$. We also note that $y=0$ correspond to step $0$ for each waveguide, and $y=nL$ correspond to step $n$. The input-output operator in the manuscript is nothing but:
\begin{equation}
\begin{aligned}
b_{n,x;L,in}(t) = b_{x-n,+}(nL-\epsilon,t)\\
b_{n,x;R,out}(t) = b_{x-n,+}(nL+\epsilon,t)\\
b_{n,x;R,in}(t) = b_{x+n,-}(nL-\epsilon,t)\\
b_{n,x;L,out}(t) = b_{x+n,-}(nL+\epsilon,t)\\
\end{aligned}
\label{eq1}
\end{equation}
where $\epsilon$ is an infinitesimal positive number.

\subsection{Hamiltonian for the system}
Firstly, the Hamiltonian for the waveguide photons are:
\begin{equation}
H_{ph} = \sum_{m,\nu}\int dk \ k\tilde{b}_{m,\nu}(k)^{\dagger}\tilde{b}_{m,\nu}(k)
\end{equation}
Here $\tilde{b}_{m,\nu}(k)$ is the Fourier transform of the operator $b_{m,\nu}(y)$.
Second, an atom located at step $n$ position $x$ have its own system Hamiltonian $H_{atom,n,x}$. 

Finally, the interaction term between the waveguide mode and an atom located at $(n,x)$ are given by:
\begin{equation}
H_{int,n,x} = V\int dk (e^{-inkL}\tilde{b}^{\dagger}_{x-n,+}(k)a_{n,x} + e^{-inkL}\tilde{b}^{\dagger}_{x+n,-}(k)a_{n,x}) + h.c.
\end{equation}
The full Hamiltonian is thus the sum of all terms above.

\subsection{Deriving input-output relations from the Hamiltonian}
First, let us consider the bosonic field operator $\tilde{b}_{m,+}(k)$. Directly commuting it with the system Hamiltonian gives:
\begin{equation}
i\frac{d}{dt'}\tilde{b}_{m,+}(k,t') = k\tilde{b}_{m,+}(k,t') + \sum_{n} Va_{n,n+m}(t')e^{-inkL}
\end{equation}
Multiplying both sides with $e^{ikt'}$ and integrate in the interval $(t_0,t)$ gives:
\begin{equation}
\tilde{b}_{m,+}(k,t)e^{ikt}-\tilde{b}_{m,+}(k,t_0)e^{ikt_0} = -iV\sum_{n} \int_{t_0}^{t}a_{n,n+m}(t')e^{-inkL}e^{ikt'}
\end{equation}

Now we use the relations: $ b_{m,\nu}(y,t)=\frac{1}{\sqrt{2\pi}}\int dk\tilde{b}_{m,\nu}(k,t)e^{iky}$. Multiply the above equations with $\frac {1}{\sqrt{2\pi}}e^{iky}$ and integrate over all $k$ gives:
\begin{equation}
b_{m,+}(y+t,t)-b_{m,+}(y+t_0,t_0) = -i\sqrt{2\pi}V\sum_{n} \int_{t_0}^{t}a_{n,n+m}(t')\delta(t'-nL+y)
\end{equation}
Note that we may choose any time interval $(t_0,t)$ and $y$ as we want. For example, we may have:
\begin{equation}
b_{m,+}(nL+\epsilon,t^*+\epsilon)-b_{m,+}(nL-\epsilon,t^*-\epsilon) = -i\sqrt{\gamma}a_{n,n+m}(t^*)
\end{equation}
by choosing $\sqrt{\gamma}=\sqrt{2\pi V^2}$, which is the input-output relations in the main manuscript.
Also note that we have $b_{m,+}(nL-\epsilon,t)=b_{m,+}((n-1)L+\epsilon,t-L)$. Generally, we have \ref{eq2} and \ref{eq3}.

For the atomic operators $a_{n,x}(t)$, directly commuting it with the Hamiltonian gives:
\begin{equation}
\begin{aligned}
\frac{da_{n,x}(t)}{dt} = -i\sqrt{\gamma}&[a_{n,x}(t),a^{\dagger}_{n,x}(t)](b_{x-n,+}(nL,t)+b_{x+n,-}(nL,t))\\
-i&[a_{n,x}(t),H_{atom,n,x}(t)]\\
\end{aligned}
\end{equation}
But we should always keep in mind that $b_{m,+}(nL,t)=\frac{1}{2}(b_{m,+}(nL+\epsilon,t)+b_{m,+}(nL-\epsilon,t))$, that is, the atom interact with the average field before and in front of it\cite{Fan2010input}. Thus, plugging in, we have:
\begin{equation}
\begin{aligned}
\frac{da_{n,x}(t)}{dt} = -i\sqrt{\gamma}&[a_{n,x}(t),a^{\dagger}_{n,x}(t)](b_{x-n,+}(nL-\epsilon,t)+b_{x+n,-}(nL-\epsilon,t)-i\sqrt{\gamma}a_{n,x})\\
-i&[a_{n,x}(t),H_{atom,n,x}(t)]\\
\end{aligned}
\end{equation}
Plugging in the formula for the input and output operators for each atom, we obtain:
\begin{equation}
\begin{aligned}
\frac{da_{n,x}(t)}{dt} = -i\sqrt{\gamma}&[a_{n,x}(t),a^{\dagger}_{n,x}(t)](b_{n,x;L,in}(t)+b_{n,x;R,in}(t)-i\sqrt{\gamma}a_{n,x})\\
-i&[a_{n,x}(t),H_{atom,n,x}(t)]\\
\end{aligned}
\end{equation}
as shown in the main manuscript.

\subsection{Summary of input-output relations}
Similar to \cite{Xu2015,Caneva2015a}, we have the following formula for $m\neq 0$:
\begin{equation}
\begin{aligned}
b_{m,+}(y>0,t) = b_{m,+}(-\epsilon,t-y) - i\sqrt{\gamma}\sum_{n}\theta(y-nL)a_{n,m+n}(t-(y-nL)) (m\neq 0)\\
b_{m,-}(y>0,t) = b_{m,-}(-\epsilon,t-y)- i\sqrt{\gamma}\sum_{n}\theta(y-nL)a_{n,m-n}(t-(y-nL)) (m\neq 0)\\
\end{aligned}
\label{eq2}
\end{equation}

For $m=0$, 
\begin{equation}
\begin{aligned}
b_{0,+}(y>0,t) = \frac{1}{\sqrt{2}}[b_{0,+}(-\epsilon,t-y)+b_{0,-}(-\epsilon,t-y)] - i\sqrt{\gamma}\sum_{n}\theta(y-nL)a_{n,n}(t-(y-nL))\\
b_{0,-}(y>0,t) = \frac{1}{\sqrt{2}}[b_{0,+}(-\epsilon,t-y)-b_{0,-}(-\epsilon,t-y)]- i\sqrt{\gamma}\sum_{n}\theta(y-nL)a_{n,-n}(t-(y-nL))\\
\end{aligned}
\label{eq3}
\end{equation}
In the above equation, $\theta(y)$ is the heaviside function. The reason why $b_{0,\pm}$ is a bit more complicated is simply because we have placed a "source" 50:50-linear-beamsplitter at step $0$, $x=0$. 

As for the dynamics of an atom, located at $(n,x)$, we have:

\begin{equation}
\begin{aligned}
\frac{da_{n,x}(t)}{dt} = -i\sqrt{\gamma}&[a_{n,x}(t),a^{\dagger}_{n,x}(t)](b_{n,x;L,in}(t)+b_{n,x;R,in}(t))\\
-i&[a_{n,x}(t),H_{atom,n,x}(t)-i\gamma a^{\dagger}_{n,x}(t) a_{n,x}(t)]\\
\end{aligned}
\end{equation}

Canonically, one should choose the atomic operator $a$ to be the lowering operator $\sigma_{-}$. But to simplify the presentation of the resulting equations, we can alternatively choose $a$ to be bosonic, as in \cite{Caneva2015a}, while introducing a 
"penalty-term" $(U_0/2)a^{\dagger} a(a^{\dagger} a - 1)$, so that the Hamiltonian for each atom becomes $H_{atom}=\Omega a^{\dagger}a + (U_0/2)a^{\dagger} a(a^{\dagger} a - 1)$. The two-level nature is captured by taking the limit $U_0\rightarrow \infty$ at the end of calculations for observable quantities.

The explicit formula for the dynamics of atomic operator $a$ for each $(n,x)$ are given by:
\begin{equation}
\begin{aligned}
\dot{a}_{n,x} =  -i\Omega a_{n,x}(t)-iU_{0}a^{\dagger}_{n,x}(t)a_{n,x}(t)a_{n,x}(t)-
\gamma\sum_{\substack{n'\leq n\\|n-n'|=|x-x'|}}a_{n',x'}(t-|n-n'|L)\\
-i\sqrt{\gamma}(b_{x-n,+}(-\epsilon,t-nL)+b_{x+n,-}(-\epsilon,t-nL))
\end{aligned}
\label{eq5}
\end{equation}
for $|x|\neq n$.

For $|x|=n$, we obtain:
\begin{equation}
\begin{aligned}
\dot{a}_{n,n} =  -i\Omega a_{n,n}(t)-iU_{0}a^{\dagger}_{n,n}(t)a_{n,n}(t)a_{n,n}(t)-
\gamma\sum_{\substack{n'\leq n}}a_{n',n'}(t-|n-n'|L)\\
-i\sqrt{\gamma}(\frac{1}{\sqrt{2}}[b_{0,+}(-\epsilon,t-nL)+b_{0,-}(-\epsilon,t-nL)]+b_{2n,-}(-\epsilon,t-nL))
\end{aligned}
\label{eq6}
\end{equation}
and
\begin{equation}
\begin{aligned}
\dot{a}_{n,-n} =  -i\Omega a_{n,-n}(t)-iU_{0}a^{\dagger}_{n,n}(t)a_{n,n}(t)a_{n,n}(t)-
\gamma\sum_{\substack{n'\leq n}}a_{n',n'}(t-|n-n'|L)\\
-i\sqrt{\gamma}(\frac{1}{\sqrt{2}}[b_{0,+}(-\epsilon,t-nL)-b_{0,-}(-\epsilon,t-nL)]+b_{-2n,+}(-\epsilon,t-nL))
\end{aligned}
\label{eq7}
\end{equation}

As shown in the main manuscript, we are injecting a two-photon pulse towards the source beamsplitter from the left. We can therefore write the state as $\ket{init_2}=\ket{init_1}\otimes\ket{init_1}$ for $\ket{init_1}=\int_{-\infty}^{0}dy\sqrt{2\kappa}b_{0,+}(y,0)e^{(\kappa+i\omega)y}\ket{vac}$\cite{Xu2015}. This represent the state at time $T=0$. At time $T\geq NL$, the photo detectors located at $y=NL+\epsilon$ will detect photons. 
For a two-photon initial state shown above, we are interested in the quantities $\bra{vac}b_{N,x_1;d_1,out}(t+\tau+NL) b_{N,x_2;d_2,out}(t+NL)\ket{init_{2}}$. This is because we have:
\begin{equation}
\begin{aligned}
&||\bra{vac}b_{N,x_1;d_1,out}(t+\tau+NL) b_{N,x_2;d_2,out}(t+NL)\ket{init_{2}}||^2\\
&=\bra{init_2}b^{\dagger}_{N,x_2;d_2,out}(t+NL)b^{\dagger}_{N,x_1;d_1,out}(t+\tau+NL)b_{N,x_1;d_1,out}(t+\tau+NL) b_{N,x_2;d_2,out}(t+NL)\ket{init_{2}}
\end{aligned}
\end{equation}
Therefore, we should compute the quantities:
$\bra{vac}b_{m_2,\nu_2}(NL+\epsilon,t+\tau+NL) b_{m_1,\nu_1}(NL+\epsilon,t+NL)\ket{init_{2}}$ for $t+\tau\geq t\geq0$.
for all possible $m\in2\mathbf{Z}$ and $\nu=\pm$.

From \ref{eq2} and \ref{eq3}, these correlations can be decomposed into correlations of the following four types:
$\bra{vac}b_{m,\nu}(-\epsilon,t+\tau) b_{m,\nu}(-\epsilon,t)\ket{init_{2}}$, $\bra{vac}a_{n,x}(t+\tau+nL) b_{m,\nu}(-\epsilon,t)\ket{init_{2}}$,
$\bra{vac}b_{m,\nu}(-\epsilon,t+\tau) a_{n,x}(t+nL)\ket{init_{2}}$ and $\bra{vac}a_{n',x'}(t+\tau+n'L) a_{n,x}(t+nL)\ket{init_{2}}$.
Here $m$ is an even integer labeling the waveguides and $\nu\in\pm$ is the pointing direction of the waveguide.

\section{Solving correlations for $L=0$}
Here we would compute the special case of $L\rightarrow 0$. Therefore, time retardation can be neglected. However, as shown in \cite{Pletyukhov2012} as well as the next section, for arbitrary $L$, the output photon correlations measured at the end of the network does not depend on the spatial separation $L$. Namely, the final output state can always be obtained by iteratively applying the unitary tranformation implemented by each step of the walk to the initial input state.

\subsection{Single-photon correlations}
As a warm-up, we start with computing single photon correlations. Namely, we shall compute $\bra{vac}b_{m,\nu}(+\epsilon,t)\ket{init_{1}}$. According to \ref{eq2} and \ref{eq3}, this quantity can be decomposed into linear combinations of $\bra{vac}-ia_{n,x}(t)\ket{init_1}$ as well as $\bra{vac}b_{m,\nu}(-\epsilon,t)\ket{init_1}$. Therefore, we only need to compute these correlations. 

Firstly, note that the correlation $\bra{vac}b_{m,\nu}(-\epsilon,t)\ket{init_1}=\bra{vac}b_{m,\nu}(-t,0)\ket{init_1}$. 
Hence, we have $\bra{vac}b_{0,+}(-\epsilon,t)\ket{init_1}=\sqrt{2\kappa}e^{-(\kappa+i\omega) t}$ and $0$ otherwise. 
Furthermore, by using equation \ref{eq5}, \ref{eq6}, \ref{eq7} we can write down the differential equation for $\bra{vac}-ia_{n,x}(t)\ket{init_1}$ as:
\begin{equation}
\begin{aligned}
\frac{d}{dt}\bra{vac}-ia_{n,x}(t)\ket{init_1} = -(\gamma+i\Omega)\bra{vac}-ia_{n,x}(t)\ket{init_1}\\
-\gamma \sum_{\substack{n'< n\\|n-n'|=|x-x'|}}\bra{vac}-ia_{n',x'}(t)\ket{init_1}\\
\end{aligned}
\end{equation}
for $|x|\neq n$. For $|x|=n$, we have:
\begin{equation}
\begin{aligned}
\frac{d}{dt}\bra{vac}-ia_{n,n}(t)\ket{init_1} = -(\gamma+i\Omega)\bra{vac}-ia_{n,n}(t)\ket{init_1}\\
-\gamma \sum_{n'<n}\bra{vac}-ia_{n',n'}(t)\ket{init_1}-\sqrt{\kappa\gamma}e^{-(\kappa+i\omega) t}
\end{aligned}
\end{equation}

We can therefore use an ansatz $\bra{vac}-ia_{n,x}(t)\ket{init_1}=(p_{1}(t)e^{-\gamma t}+p_{2}(t)e^{-(\kappa+i\delta) t})e^{-i\Omega t}$ where $p_{i}(t)$ are $(n,x)$-dependent polynomials of $t$ and $\delta\equiv \omega-\Omega$. We then recursively find the correlation functions, starting from $(n,x)=(1,1)$ and $(n,x)=(1,-1)$
In fact, these ansatz apply to broader set of correlations: $\bra{vac}a_{n',x'}(t)a^{\dagger}_{n,x}(0)\ket{vac}$. One should note that this quantity is only non-zero for $(n',x')$ in the light-cone of $(n,x)$, namely $n'\geq n$ and $|x-x'|\leq|n-n'|$.
The differential equations for $\bra{vac}a_{n',x'}(t)a^{\dagger}_{n,x}(0)\ket{vac}$ is similarly given by:
\begin{equation}
\begin{aligned}
&\frac{d}{dt}\bra{vac}a_{n',x'}(t)a^{\dagger}_{n,x}(0)\ket{vac} \\
&= -\gamma \sum_{\substack{n''< n'\\|n''-n'|=|x''-x'|}}\bra{vac}a_{n'',x''}(t)a^{\dagger}_{n,x}(0)\ket{vac}\\
&-(\gamma+i\Omega)\bra{vac}a_{n',x'}(t)a^{\dagger}_{n,x}(0)\ket{vac}
\end{aligned}
\end{equation}

Note that we have ignored terms such as $\bra{vac}b_{m,\nu}(-\epsilon,t)a^{\dagger}_{n,x}(0)\ket{vac}$ since they are zero from causality. 
Also, one can easily verify that $\bra{vac}a_{n,x}(t)a^{\dagger}_{n,x}(0)\ket{vac}=e^{-(\gamma+i\Omega) t}$. Then we can apply the ansatz $\bra{vac}a_{n',x'}(t)a^{\dagger}_{n,x}(0)\ket{vac}=(p_{3}(t)e^{-\gamma t})e^{-i\Omega t}$ and find $\bra{vac}a_{n',x'}(t)a^{\dagger}_{n,x}(0)\ket{vac}$ recursively, starting from $(n',x')=(n+1,x-1)$ and $(n',x')=(n+1,x+1)$.

\subsection{Two-photon correlations}
Now we would compute two-photon correlations. To begin with, we shall first focus on three types of simpler correlations 
$\bra{vac}b_{m',\nu'}(y<0,t)b_{m,\nu}(-\epsilon,t)\ket{init_{2}}$, $\bra{vac}(-ia_{n,x}(t))b_{m,\nu}(y<0,t)\ket{init_{2}}$ 
and $\bra{vac}(-ia_{n',x'}(t))(-ia_{n,x}(t))\ket{init_{2}}$. We now consider them separately.

\subsubsection{Case 1}
Again, we start from the simplest case
\begin{equation}
\bra{vac}b_{m',\nu'}(y<0,t)b_{m,\nu}(-\epsilon,t)\ket{init_{2}}=
\bra{vac}b_{m',\nu'}(-\epsilon,t)b_{m,\nu}(-\epsilon,t)\ket{init_{2}}e^{(\kappa+i\omega) y}
=2\sqrt{2}\kappa e^{-2(\kappa+i\omega) t}e^{(\kappa+i\omega) y}
\end{equation}
if and only if $(m,\nu)=(m',\nu')=(0,+)$. As a special case, we have:
$\bra{vac}b_{0,+}(-\epsilon,t)b_{0,+}(-\epsilon,t)\ket{init_{2}}=2\sqrt{2}\kappa e^{-2(\kappa+i\omega) t}=2\sqrt{2}\kappa e^{-2(\kappa+i\delta) t}e^{-2i\Omega t}$.

\subsubsection{Case 2}
For the case of $\bra{vac}(-ia_{n,x}(t))b_{m,\nu}(y<0,t)\ket{init_{2}}$, we derive them in a similar fashion.

First note that, only for $(m,\nu)=(0,+)$:
\begin{equation}
\begin{aligned}
\bra{vac}(-ia_{n,x}(t))b_{m,\nu}(y<0,t)\ket{init_{2}}&=\bra{vac}(-ia_{n,x}(t))\ket{init_{1}}\sqrt{2}\sqrt{2\kappa}e^{(\kappa+i\omega) (y-t)}\\
&=\bra{vac}(-ia_{n,x}(t))b_{m,\nu}(-\epsilon,t)\ket{init_{2}}e^{(\kappa+i\omega) y}
\end{aligned}
\end{equation}

But we have already derived $\bra{vac}(-i)a_{n,x}(t)\ket{init_{1}}$ in the previous section.
We therefore know that 
$\bra{vac}(-i)a_{n,x}(t)b_{m,\nu}(-\epsilon,t)\ket{init_{2}}$ has the form of 
$(q_{1}(t)e^{-2(\kappa+i\delta) t}+q_{2}(t)e^{-(\kappa+i\delta+\gamma)t})e^{-i2\Omega t}$.

\subsubsection{Case 3}
Finally, we need to deal with $\bra{vac}a_{n',x'}(t)a_{n,x}(t)\ket{init_{2}}$. From \ref{eq5}, \ref{eq6}, \ref{eq7}, we know that, for $(n',x')\neq(n,x)$ 
satisfying $|x|\neq n$ and  $|x'|\neq n'$ :
\begin{equation}
\begin{aligned}
&\frac{d}{dt}\bra{vac}(-ia_{n',x'}(t))(-ia_{n,x}(t))\ket{init_2} = \\
&-(2i\Omega+2\gamma)\bra{vac}(-ia_{n',x'}(t))(-ia_{n,x}(t))\ket{init_2}\\
&-\gamma \sum_{\substack{n''< n\\|n-n''|=|x-x''|}}\bra{vac}(-ia_{n',x'}(t))(-ia_{n'',x''}(t))\ket{init_2}\\
&-\gamma \sum_{\substack{n'''< n'\\|n-n'''|=|x-x'''|}}\bra{vac}(-ia_{n''',x'''}(t))(-ia_{n,x}(t))\ket{init_2}\\
\end{aligned}
\end{equation}
after throwing away terms such as $-\sqrt{\gamma}\bra{vac}(-ia_{n',x'}(t))(b_{x-n,+}(-\epsilon,t)+b_{x+n,-}(-\epsilon,t))\ket{init_{2}}$ because they are $0$ for $|x|\neq n$. For $x=n$, we have:
\begin{equation}
\begin{aligned}
&\frac{d}{dt}\bra{vac}(-ia_{n',x'}(t))(-ia_{n,x}(t))\ket{init_2} = \\
&-(2i\Omega+2\gamma)\bra{vac}(-ia_{n',x'}(t))(-ia_{n,x}(t))\ket{init_2}\\
&-\gamma \sum_{\substack{n''< n\\|n-n''|=|x-x''|}}\bra{vac}(-ia_{n',x'}(t))(-ia_{n'',x''}(t))\ket{init_2}\\
&-\gamma \sum_{\substack{n'''< n'\\|n-n'''|=|x-x'''|}}\bra{vac}(-ia_{n''',x'''}(t))(-ia_{n,x}(t))\ket{init_2}\\
&-\sqrt{\gamma}\bra{vac}(-ia_{n',x'}(t))(\frac{1}{\sqrt{2}}b_{0,+}(-\epsilon,t)\ket{init_{2}}\\
\end{aligned}
\end{equation}

But we have already obtained all the terms on the last line from \textit{Case 2}. We can therefore use the ansatz
$\bra{vac}a_{n',x'}(t)a_{n,x}(t)\ket{init_{2}}=(q_{1}(t)e^{-2(\kappa+i\delta) t}+q_{2}(t)e^{-(\kappa+i\delta+\gamma)t}+q_{3}(t)e^{-2\gamma t})e^{-2i\Omega t}$
and recursively find the $(n,x),(n',x')$ dependent polynomials $q_{1}(t),q_{2}(t),q_{3}(t)$.

We also claim that $\bra{vac}a_{n,x}(t)a_{n,x}(t)\ket{init_{2}}\rightarrow 0$. Again, from \ref{eq5}, \ref{eq6}, \ref{eq7},
\begin{equation}
\begin{aligned}
\frac{d}{dt}\bra{vac}(-ia_{n,x}(t))(-ia_{n,x}(t))\ket{init_2} &=\bra{vac}(-ia_{n,x}(t))(-U_{0}a^{\dagger}_{n,x}(t)a_{n,x}(t)a_{n,x}(t))\ket{init_{2}}+(\cdots)\\
&=iU_{0}\bra{vac}(a_{n,x}(t)a_{n,x}(t))\ket{init_{2}}+(\cdots)
\end{aligned}
\end{equation}
Therefore, we have:
\begin{equation}
e^{iU_{0}t}\bra{vac}(a_{n,x}(t))(a_{n,x}(t))\ket{init_2} = \int_{0}^{t}F(t)e^{iU_{0}t}dt
\end{equation}
As $U_{0}\rightarrow \infty$, the integral on the right hand side converges to $0$ since $F(t)$ is a bounded continuous function.

\subsection{Final result}
In this part, we show that all the correlations that we want to compute are just the sum of products of single-photon correlations and two-photon correlations as computed above. The final correlations is just a linear combination of the following four types of correlations:
$\bra{vac}b_{m',\nu'}(0,t+\tau)b_{m,\nu}(0,t)\ket{init_{2}}$,$\bra{vac}(-ia_{n,x}(t+\tau))b_{m,\nu}(0,t)\ket{init_{2}}$,
$\bra{vac}b_{m,\nu}(0,t+\tau)(-ia_{n,x}(t))\ket{init_{2}}$,$\bra{vac}(-ia_{n',x'}(t+\tau))(-ia_{n,x}(t))\ket{init_{2}}$.
We also note that we can insert the single photon projector between the two operators, defined by:
\begin{equation}
\begin{aligned}
\Pi = \sum_{m,\nu}\int_{-\infty}^{\infty} dy b^{\dagger}_{m,\nu}(y,t)\ket{vac}\bra{vac}b_{m,\nu}(y,t) + \sum_{n,x}a^{\dagger}_{n,x}(t)\ket{vac}\bra{vac}a_{n,x}(t)=\Pi_{1}+\Pi_{2}
\end{aligned}
\end{equation}
We note that: 
$\bra{vac}(-ia_{n,x}(t+\tau))b_{m,\nu}(-\epsilon,t)\ket{init_{2}}=\bra{vac}(-ia_{n,x}(t+\tau))\Pi b_{m,\nu}(-\epsilon,t)\ket{init_{2}}$
and similarly for all other correlations.

We therefore consider these case separately:

\subsubsection{Case 1: $\Pi_{1}$}
In this section we would focus on correlations such as $\bra{vac}(O_{2}(t+\tau))\Pi_{1}(O_{1}(t))\ket{init_{2}}$.
For example, consider $\bra{vac}(-ia_{n,x}(t+\tau))b_{m,\nu}(-\epsilon,t)\ket{init_{2}}$, we have:
\begin{equation}
\begin{aligned}
&\bra{vac}(-ia_{n,x}(t+\tau))\Pi_{1}b_{m,\nu}(-\epsilon,t)\ket{init_{2}}\\
&=\bra{vac}(-ia_{n,x}(t+\tau))\sum_{m',\nu'}
\int_{-\infty}^{\infty} dy b^{\dagger}_{m',\nu'}(y,t)\ket{vac}\bra{vac}b_{m',\nu'}(y,t)b_{m,\nu}(-\epsilon,t)\ket{init_{2}}\\
&=\sum_{m',\nu'}\bra{vac}(-ia_{n,x}(t+\tau))\int_{-\infty}^{0} dy e^{(\kappa+i\omega) y}b^{\dagger}_{m',\nu'}(y,t)\ket{vac}\bra{vac}b_{m',\nu'}(-\epsilon,t)b_{m,\nu}(-\epsilon,t)\ket{init_{2}}\\
\end{aligned}
\end{equation}

Note that we are only integrating over $y\in(-\infty,0)$ since $\bra{vac}(-ia_{n,x}(t+\tau))b^{\dagger}_{m',\nu'}(y,t)\ket{vac}$ is non-zero only for $y<0$ due to causality. Namely, any excitation at $y>0$ at time $t$ cannot excite the atoms since the atoms at located around $y=0$ and the waveguides are unidirectional.

But now $\bra{vac}b_{m',\nu'}(-\epsilon,t)b_{m,\nu}(-\epsilon,t)\ket{init_{2}}$ is only non-zero for $(m,\nu)=(m',\nu')=(0,+)$ and has been computed in the previous section. Therefore, we only need to compute:
\begin{equation}
\begin{aligned}
&\bra{vac}(-ia_{n,x}(t+\tau))\int_{-\infty}^{0} dy e^{(\kappa+i\omega) y}b^{\dagger}_{0,+}(y,t)\ket{vac}2\sqrt{2}\kappa e^{-2(\kappa+i\omega) t}\\
&=\sqrt{2}\bra{vac}(-ia_{n,x}(\tau))\ket{init_{1}}\sqrt{2\kappa}e^{-2(\kappa+i\omega) t}\\
\end{aligned}
\end{equation}

Where the last line is due the fact that the Hamiltonian is time-independent and the correlations are time-translational invariant.
But we have computed $\bra{vac}(-ia_{n,x}(\tau))\ket{init_{1}}$ in the previous section as single-photon correlations.
For other correlations such as $\bra{vac}(-ia_{n',x'}(t+\tau))(-ia_{n,x}(t))\ket{init_{2}}$, we repeat the procedure as above.
Also, note that we have obtain $\bra{vac}b_{m',\nu'}(-\epsilon,t+\tau)b_{m,\nu}(-\epsilon,t)\ket{init_{2}}$ for free, given by: 
$\bra{vac}b_{m',\nu'}(-\epsilon,t+\tau)b_{m,\nu}(-\epsilon,t)\ket{init_{2}}=\sqrt{2}(2\kappa)e^{-2(\kappa+i\omega) t-(\kappa+i\omega) \tau}$
only for $(m,\nu)=(m',\nu')=(0,+)$.

\subsubsection{Case 2: $\Pi_{2}$}
In this case, we focus on $\bra{vac}(O_{2}(t+\tau))\Pi_{2}(O_{1}(t))\ket{init_{2}}$.
We note that only non-zero terms are: $\bra{vac}(-ia_{n,x}(t+\tau))\Pi_{2}b_{m,\nu}(-\epsilon,t)\ket{init_{2}}$
and $\bra{vac}(-ia_{n',x'}(t+\tau))\Pi_{2}(-ia_{n,x}(t))\ket{init_{2}}$. Plugging in 
$\Pi_{2}=\sum_{n,x}a^{\dagger}_{n,x}(t)\ket{vac}\bra{vac}a_{n,x}(t)$, we see that:
\begin{equation}
\begin{aligned}
&\bra{vac}(-ia_{n,x}(t+\tau))\Pi_{2}b_{m,\nu}(-\epsilon,t)\ket{init_{2}}\\
&=\sum_{n',x'}\bra{vac}(-ia_{n,x}(t+\tau))a^{\dagger}_{n',x'}(t)\ket{vac}\bra{vac}a_{n',x'}(t)b_{m,\nu}(-\epsilon,t)\ket{init_{2}}
\end{aligned}
\end{equation}
But each term in the sum is a product of single-photon correlations and two-photon correlations we have just computed in the previous section.
Similarly, we can compute $\bra{vac}(-ia_{n',x'}(t+\tau))\Pi_{2}(-ia_{n,x}(t))\ket{init_{2}}$ as a sum of product of single and 
two-photon correlations.

\section{Generalizing to arbitrary step separation $L$}
Here we would use some tricks to show that $L$ does not affect the final correlations measured by the detectors. 
Note that this is actually a standard result as derived in \cite{Pletyukhov2012}. But we shall re-derive the same arguement on our own.

Before we begin, we would use the notation $\tilde{a}_{n,x}(t+nL)$ to denote the operators in the general model with step-separation $L$. 
We would still use $a_{n,x}(t)$ to denote the operators in the special model of $L\rightarrow0$. Note that the initial state $\ket{init_{1}}$ 
and $\ket{init_{2}}$ are defined in the same way. From \ref{eq2}, \ref{eq3}, our fundamental building blocks are:
$\bra{vac}b_{m',\nu'}(-\epsilon,t+\tau)b_{m,\nu}(-\epsilon,t)\ket{init_{2}}$,
$\bra{vac}(-i\tilde{a}_{n,x}(t+\tau+nL))b_{m,\nu}(-\epsilon,t)\ket{init_{2}}$,
$\bra{vac}b_{m,\nu}(-\epsilon,t+\tau)(-i\tilde{a}_{n,x}(t+nL))\ket{init_{2}}$,
$\bra{vac}(-i\tilde{a}_{n',x'}(t+\tau+n'L))(-i\tilde{a}_{n,x}(t+nL))\ket{init_{2}}$.

Our goal is to show that $\bra{vac}(-i\tilde{a}_{n',x'}(t+\tau+n'L))(-i\tilde{a}_{n,x}(t+nL))\ket{init_{2}}=
\bra{vac}(-ia_{n',x'}(t+\tau))(-ia_{n,x}(t))\ket{init_{2}}$ and similarly for other correlation functions.

\subsection{Single-photon correlations}
We will again start with the simple case of $\bra{vac}(-i\tilde{a}_{n,x}(t+nL))\ket{init_1}$ as well as 
$\bra{vac}b_{m,\nu}(-\epsilon,t)\ket{init_1}$.
We have 
\begin{equation}
\begin{aligned}
\frac{d}{dt}\bra{vac}-i\tilde{a}_{n,x}(t+nL)\ket{init_1} = -\gamma \sum_{\substack{n'\leq n\\|n-n'|=|x-x'|}}
\bra{vac}-i\tilde{a}_{n',x'}(t+n'L)\ket{init_1}\\
-\sqrt{\gamma}\bra{vac}(b_{x-n,+}(-\epsilon,t)+b_{x+n,-}(-\epsilon,t))\ket{init_{1}}
\end{aligned}
\end{equation}
Also, note that the initial condition is $\bra{vac}(-i\tilde{a}_{n,x}(nL))\ket{init_1}=0$. Therefore, we can recursively show that 
$\bra{vac}(-i\tilde{a}_{n,x}(t+nL))\ket{init_1}=\bra{vac}(-ia_{n,x}(t))\ket{init_1}$ for all $n$. To be precise, we can write down the differential equations for $(\bra{vac}(-i\tilde{a}_{n,x}(t+nL))\ket{init_1}-\bra{vac}(-ia_{n,x}(t))\ket{init_1})$ and show that they are $0$, starting from $(n,x)=(1,-1)$.
In a similar way, one can show that $\bra{vac}\tilde{a}_{n',x'}(t+n'L)\tilde{a}^{\dagger}_{n,x}(nL)\ket{vac}=\bra{vac}a_{n',x'}(t)a^{\dagger}_{n,x}(0)\ket{vac}$

\subsection{Two-photon correlations}
Unlike the $L\rightarrow0$ case, now we need to take care of the ordering issue. 
Namely, we do not necessarily have  
$\tilde{a}_{n',x'}(t+n'L)\tilde{a}_{n,x}(t+nL)=\tilde{a}_{n,x}(t+nL)\tilde{a}_{n',x'}(t+n'L)$, but we would show that we still have 
$\bra{vac}\tilde{a}_{n',x'}(t+n'L)\tilde{a}_{n,x}(t+nL)\ket{init_2}=\bra{vac}\tilde{a}_{n,x}(t+nL)\tilde{a}_{n',x'}(t+n'L)\ket{init_2}=
\bra{vac}a_{n',x'}(t)a_{n,x}(t)\ket{init_2}$. We would proceed in two steps:

\subsubsection{step 1}
Firstly, we would show that $\bra{vac}\tilde{a}_{n,x}(t+nL)b_{m,\nu}(y<0,t)\ket{init_2}=\bra{vac}a_{n,x}(t)b_{m,\nu}(y<0,t)\ket{init_2}$.
Again, this is obvious since:
\begin{equation}
\begin{aligned}
\bra{vac}(-i\tilde{a}_{n,x}(t+nL))b_{m,\nu}(y<0,t)\ket{init_{2}}&=\bra{vac}(-i\tilde{a}_{n,x}(t+nL))\ket{init_{1}}\sqrt{2}\sqrt{2\kappa}
e^{(\kappa + i\omega)(y-t)}\\
&=\bra{vac}(-i\tilde{a}_{n,x}(t+nL))b_{m,\nu}(y<0,t)\ket{init_{2}}e^{(\kappa+i\omega) y}
\end{aligned}
\end{equation}
only for $(m,\nu)=(0,+)$. But again, from single-photon correlation computations in the previous section, we see that 
$\bra{vac}(-i\tilde{a}_{n,x}(t+nL))\ket{init_{1}}=\bra{vac}(-ia_{n,x}(t))\ket{init_{1}}$. Therefore, we have  $\bra{vac}\tilde{a}_{n,x}(t+nL)b_{m,\nu}(y<0,t)\ket{init_2}=\bra{vac}a_{n,x}(t)b_{m,\nu}(y<0,t)\ket{init_2}$.

Now we shall prove that $\bra{vac}b_{m,\nu}(y<0,t)\tilde{a}_{n,x}(t+nL)\ket{init_2}=\bra{vac}a_{n,x}(t)b_{m,\nu}(y<0,t)\ket{init_2}$, which is less obvious. First, note that $\bra{vac}b_{m,\nu}(y<0,t)\tilde{a}_{n,x}(nL)\ket{init_2}=0$. This is again because $\ket{init_2}$ takes at least $T=nL$ to propagate to 
step $n$. Now consider $\bra{vac}b_{m,\nu}(y<0,t)\tilde{a}_{n,x}(t'+nL)\ket{init_2}$ for $t'<t$. We now have(for $|x|\neq n$):
\begin{equation}
\begin{aligned}
\frac{\partial}{\partial t'}&\bra{vac}b_{m,\nu}(y<0,t)(-i\tilde{a}_{n,x}(nL+t'))\ket{init_2}\\
=&\bra{vac}b_{m,\nu}(y<0,t)(-U_{0}\tilde{a}^{\dagger}_{n,x}(nL+t')\tilde{a}_{n,x}(nL+t')\tilde{a}_{n,x}(nL+t'))\ket{init_2}\\
-&\gamma\sum_{\substack{n'< n\\|n-n'|=|x-x'|}}\bra{vac}b_{m,\nu}(y<0,t)(-i\tilde{a}_{n',x'}(t'+n'L))\ket{init_2}\\
-&(\gamma+i\omega)\bra{vac}b_{m,\nu}(y<0,t)(-i\tilde{a}_{n,x}(t'+nL))\ket{init_2}\\
\end{aligned}
\end{equation}
We claim that the first term $\bra{vac}b_{m,\nu}(y<0,t)(-U_{0}\tilde{a}^{\dagger}_{n,x}(nL+t')\tilde{a}_{n,x}(nL+t')\tilde{a}_{n,x}(nL+t'))\ket{init_2}=0$
for all $t'<t$. This is becuase
\begin{equation}
\begin{aligned}
&\bra{vac}b_{m,\nu}(y<0,t)(\tilde{a}^{\dagger}_{n,x}(nL+t')\tilde{a}_{n,x}(nL+t')\tilde{a}_{n,x}(nL+t'))\ket{init_2}\\
=&\bra{vac}b_{m,\nu}(y<0,t)\tilde{a}^{\dagger}_{n,x}(nL+t')\ket{vac}\bra{vac}\tilde{a}_{n,x}(nL+t')\tilde{a}_{n,x}(nL+t'))\ket{init_2}\\
=&(\bra{vac}\tilde{a}_{n,x}(nL+t')b^{\dagger}_{m,\nu}(y<0,t)\ket{vac})^{\dagger}
\bra{vac}\tilde{a}_{n,x}(nL+t')\tilde{a}_{n,x}(nL+t'))\ket{init_2}\\
\end{aligned}
\end{equation} 
But the term $\bra{vac}\tilde{a}_{n,x}(nL+t')b^{\dagger}_{m,\nu}(y<0,t)\ket{vac}=0$ because $t'<t$. This is simply because it takes a time 
interval $|y|+nL$ for any excitation to propagate from $y<0$ to an atom at $nL$, but the time interval for the above correlation function
$(nL+t'-t)<nL$. Therefore, we have:
\begin{equation}
\begin{aligned}
\frac{\partial}{\partial t'}&\bra{vac}b_{m,\nu}(y<0,t)(-i\tilde{a}_{n,x}(nL+t'))\ket{init_2}\\
=-&(\gamma+i\Omega)\bra{vac}b_{m,\nu}(y<0,t)(-i\tilde{a}_{n,x}(nL+t'))\ket{init_2}\\
-&\gamma\sum_{\substack{n'< n\\|n-n'|=|x-x'|}}\bra{vac}b_{m,\nu}(y<0,t)(-i\tilde{a}_{n',x'}(t'+n'L))\ket{init_2}\\
\end{aligned}
\end{equation}

For $|x|=n$, the equation is similar. It is given by:
\begin{equation}
\begin{aligned}
\frac{\partial}{\partial t'}&\bra{vac}b_{m,\nu}(y<0,t)(-i\tilde{a}_{n,n}(nL+t'))\ket{init_2}\\
=-&(\gamma+i\Omega)\bra{vac}b_{m,\nu}(y<0,t)(-i\tilde{a}_{n,n}(nL+t'))\ket{init_2}\\
-&\gamma\sum_{n'<n}\bra{vac}b_{m,\nu}(y<0,t)(-i\tilde{a}_{n',n'}(t'+n'L))\ket{init_2}\\
-&\sqrt{\gamma}\bra{vac}b_{m,\nu}(y<0,t)(\frac{1}{\sqrt{2}}b_{0,+}(-\epsilon,t'))\ket{init_2}
\end{aligned}
\end{equation}

Now consider the family of functions:
$\bra{vac}b_{m,\nu}(y<0,t)(-i\tilde{a}_{n,x}(nL+t'))\ket{init_2}-\bra{vac}b_{m,\nu}(y<0,t)(-ia_{n,x}(t'))\ket{init_2}$.

We have, for $|x|\neq n$:
\begin{equation}
\begin{aligned}
\frac{\partial}{\partial t'}&(\bra{vac}b_{m,\nu}(y<0,t)(-i\tilde{a}_{n,x}(nL+t'))\ket{init_2}-
\bra{vac}b_{m,\nu}(y<0,t)(-ia_{n,x}(t'))\ket{init_2})\\
&=-\gamma\sum_{\substack{n'< n\\|n-n'|=|x-x'|}}
(\bra{vac}b_{m,\nu}(y<0,t)(-i\tilde{a}_{n',x'}(t'+n'L))\ket{init_2}-\bra{vac}b_{m,\nu}(y<0,t)(-ia_{n',x'}(t'))\ket{init_2})\\
&-(\gamma+i\Omega)(\bra{vac}b_{m,\nu}(y<0,t)(-i\tilde{a}_{n,x}(nL+t'))\ket{init_2}-
\bra{vac}b_{m,\nu}(y<0,t)(-ia_{n,x}(t'))\ket{init_2})
\end{aligned}
\end{equation}
And we can use the fact that 
$\bra{vac}b_{m,\nu}(y<0,t)(-i\tilde{a}_{n,x}(nL))\ket{init_2}-\bra{vac}b_{m,\nu}(y<0,t)(-ia_{n,x}(0))\ket{init_2}=0$ to show that the functions defined above are all $0$ recursively, starting from $(n,x)=(1,-1)$.

Finally, we shall take $t'\rightarrow t$ and we get
$\bra{vac}b_{m,\nu}(y<0,t)\tilde{a}_{n,x}(t+nL)\ket{init_2}=\bra{vac}a_{n,x}(t)b_{m,\nu}(y<0,t)\ket{init_2}$ as desired.

\subsubsection{step 2}
Now we can prove that $\bra{vac}(-i\tilde{a}_{n',x'}(t+n'L))(-i\tilde{a}_{n,x}(t+nL))\ket{init_{2}}-
\bra{vac}(-ia_{n',x'}(t))(-ia_{n,x}(t))\ket{init_{2}}=0$. We can simply take the derivative of these functions with respect to $t$ according to
\ref{eq5}, \ref{eq6}, \ref{eq7} and show that they are $0$ recursively, starting with $n'=n=1$. 
The initial condition is given by $\bra{vac}(-i\tilde{a}_{n',x'}(n'L))(-i\tilde{a}_{n,x}(nL))\ket{init_{2}}-
\bra{vac}(-ia_{n',x'}(0))(-ia_{n,x}(0))\ket{init_{2}}=0$.


The whole idea behind this is that the family
of correlations $\bra{vac}(-i\tilde{a}_{n',x'}(t+n'L))(-i\tilde{a}_{n,x}(t+nL))\ket{init_{2}}$ and $\bra{vac}(-ia_{n',x'}(t))(-ia_{n,x}(t))\ket{init_{2}}=0$ satisfy the exact same differential equations and have the same initial condition,
therefore they have to be equal.

\subsection{Final results}
Now we can finally prove that $\bra{vac}(-i\tilde{a}_{n',x'}(t+\tau+n'L))(-i\tilde{a}_{n,x}(t+nL))\ket{init_{2}}=
\bra{vac}(-ia_{n',x'}(t+\tau))(-ia_{n,x}(t))\ket{init_{2}}$ and similarly for other correlations. We shall take $\bra{vac}(-i\tilde{a}_{n',x'}(t+\tau+n'L))(-i\tilde{a}_{n,x}(t+nL))\ket{init_{2}}$ as an example.
Consider the functions $\bra{vac}(-i\tilde{a}_{n',x'}(t+\tau+n'L))(-i\tilde{a}_{n,x}(t+nL))\ket{init_{2}}-
\bra{vac}(-ia_{n',x'}(t+\tau))(-ia_{n,x}(t))\ket{init_{2}}$. 


We now have the initial condition $\bra{vac}(-i\tilde{a}_{n',x'}(t+n'L))(-i\tilde{a}_{n,x}(t+nL))\ket{init_{2}}-
\bra{vac}(-ia_{n',x'}(t))(-ia_{n,x}(t))\ket{init_{2}}=0$, as shown in the previous section in two steps. Hence, we can take the partial derivative with respect to $\tau$ for the function $\bra{vac}(-i\tilde{a}_{n',x'}(t+\tau+n'L))(-i\tilde{a}_{n,x}(t+nL))\ket{init_{2}}-
\bra{vac}(-ia_{n',x'}(t+\tau))(-ia_{n,x}(t))\ket{init_{2}}=0$ for some fixed $t$, starting from $n=n'=1$. From these equation we can show, recursively, that these functions are all $0$.

\section{Additional numerical results}
For this section, we present the numerical results mentioned in the manuscript.

\subsection{Case 1: $\gamma=\delta$}
\begin{figure}[hbt]
	\subfloat[(a)]{%
		\includegraphics[height=4cm,width=.3\linewidth]{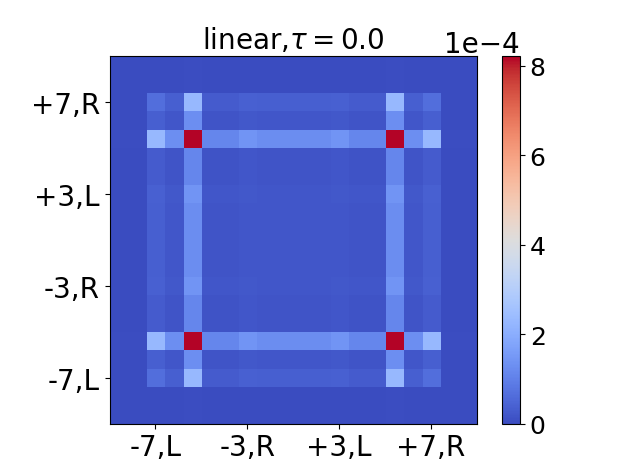}%
	}
	\subfloat[(b)]{%
		\includegraphics[height=4cm,width=.3\linewidth]{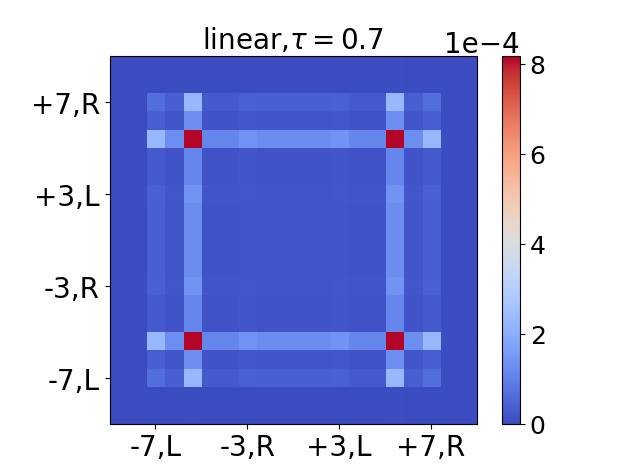}%
	}
	\subfloat[(c)]{%
		\includegraphics[height=4cm,width=.3\linewidth]{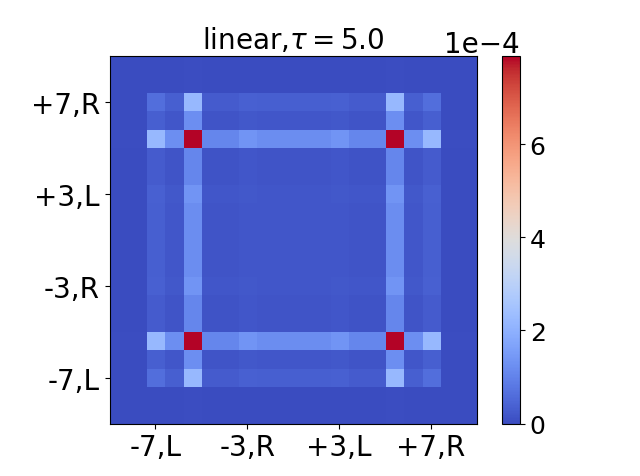}%
	}
	\caption{$\tau$-invariant statistical pattern of the linear quantum walk, for $\delta=\gamma$.}
	\label{fig:s4}
\end{figure}

Here we present further results for $\kappa\ll\gamma=\delta=\omega-\Omega$. In particular, we present the statistical patterns for the reference model (i.e. the non-interacting quantum walk with linear beam splitters). Indeed, as shown in Figure~\ref{fig:s4}, the statistical pattern is $\tau$-invariant.

\subsection{Case 2: $\delta=0$}
\begin{figure}[hbt]
	\subfloat[(a)]{%
		\includegraphics[height=4cm,width=.3\linewidth]{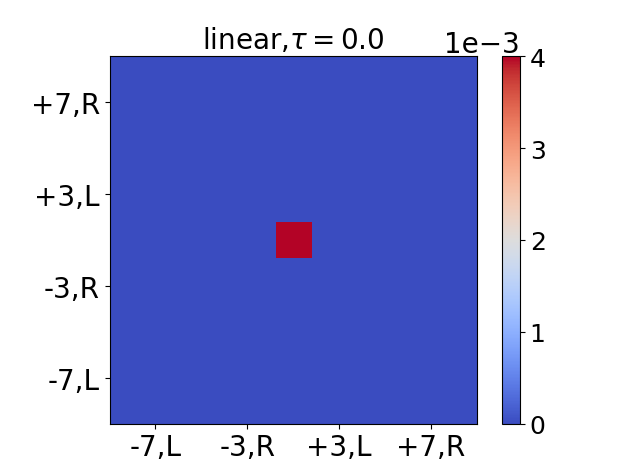}%
	}
	\subfloat[(b)]{%
		\includegraphics[height=4cm,width=.3\linewidth]{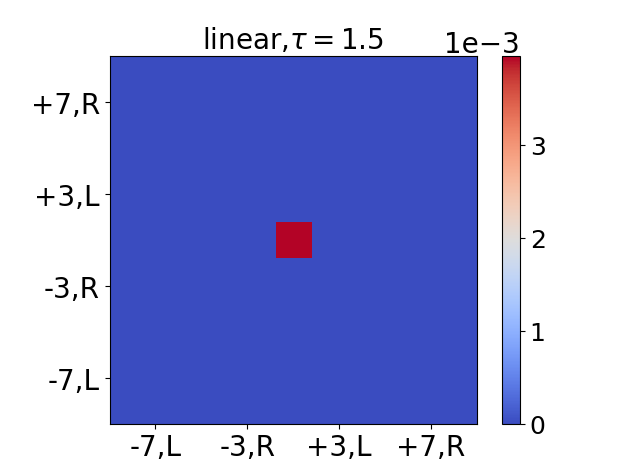}%
	}
	\subfloat[(c)]{%
		\includegraphics[height=4cm,width=.3\linewidth]{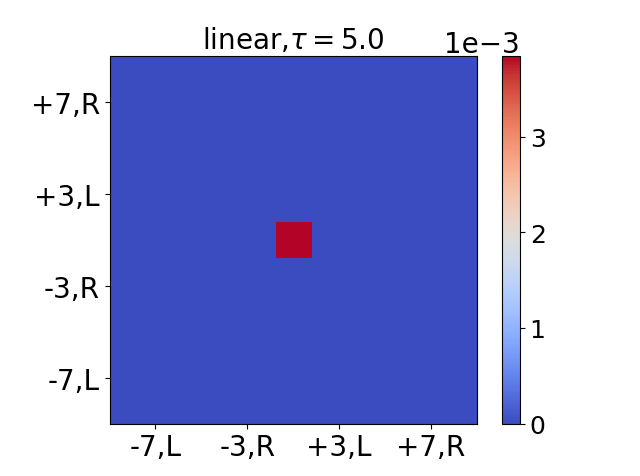}%
	}
	\caption{$\tau$-invariant statistical pattern of the linear quantum walk, for $\delta=0$.}
	\label{fig:s5}
\end{figure}

\begin{figure}[hbt]
	\subfloat[(a)$(-3,L;+3,R);(+3,R;-3,L)$]{%
		\includegraphics[height=5cm,width=.5\linewidth]{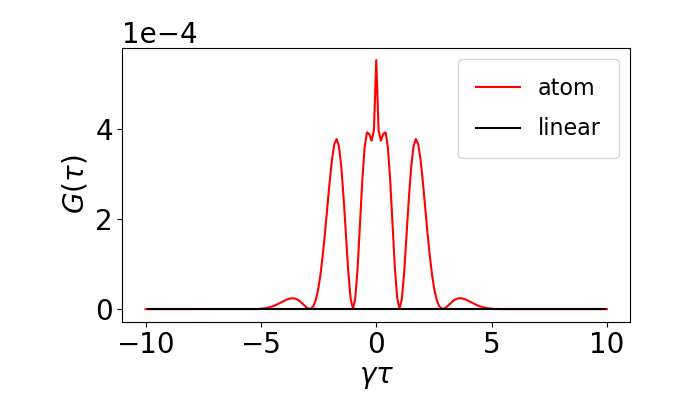}%
	}
	\subfloat[(b)$(-1,L;+1,R);(+1,R;-1,L)$]{%
		\includegraphics[height=5cm,width=.5\linewidth]{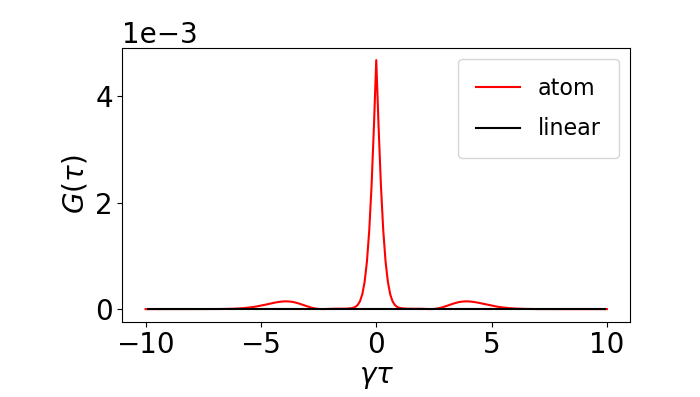}%
	}
	\caption{Interval-time correlation functions $G(\tau)$ for two pairs of output channels on the anti-correlated diagonal. Note that the nature of correlation is similar to that of the channel pair $(-5,L;+5,R);(+5,R;-5,L)$ shown in the main manuscript.}
	\label{fig:s6}
\end{figure}

For the resonant case($\delta=0$), we see once again that the statistical pattern for the linear beam splitter quantum walk is $\tau$-invariant(Figure~\ref{fig:s5}). Furthermore, we pick two extra pairs of channels on the anti-correlated diagonal $(-3,L;+3,R);(+3,R;-3,L)$ and $(-1,L;+1,R);(+1,R;-1,L)$ and compare the interval-time correlation function $G(\tau)$ for all $\tau$ for both the linear and nonlinear quantum walk(Figure \ref{fig:s6}). These two pairs of channels exhibits similar correlation functions to that of the channel pair $(-5,L;+5,R);(+5,R;-5,L)$.

\newpage

\bibliography{first_paper_ref}